\documentclass[letter, conference]{IEEEtran}
\IEEEoverridecommandlockouts

\usepackage{cite}
\usepackage{amsmath,amssymb,amsfonts}
\usepackage{algorithmic}
\usepackage{graphicx}
\usepackage{textcomp}
\usepackage{xcolor}

\usepackage{subcaption}

\usepackage{tikz, pgfplots}
\pgfplotsset{compat=1.18}
\usepgfplotslibrary{statistics}

\newcommand{\tcb}{\textcolor{black}}

\usepackage[compact]{titlesec}
\titlespacing{\section}{0pt}{2ex}{1ex}
\titlespacing{\subsection}{0pt}{1ex}{0ex}
\titlespacing{\subsubsection}{0pt}{0.5ex}{0ex}

\begin{document}

\title{Multipolar dynamics of social segregation: \\
{\LARGE Data validation on Swedish vaccination statistics}
\thanks{The authors are with the Department of Automatic Control, Lund University, Lund, Sweden. They are members of the ELLIIT Strategic Research Area at Lund University. This work is partially funded by Wallenberg AI, Autonomous Systems and Software Program (WASP) funded by the Knut and Alice Wallenberg Foundation. Email: {\tt\small\{luka.bakovic,david.ohlin,emma.tegling\} @control.lth.se}. }
}
\author{{Luka Baković, David Ohlin and Emma Tegling}}

\maketitle

\begin{abstract}
We perform a validation analysis on the multipolar model of opinion dynamics. A general methodology for using the model on datasets of two correlated variables is proposed and tested using data on the relationship between \mbox{COVID-19} vaccination rates and political participation in Sweden. The model is shown to successfully capture the opinion segregation demonstrated by the data and spatial correlation of biases is demonstrated as necessary for the result.\tcb{ A mixing of the biases on the other hand leads to a more homogeneous opinion distribution, and greater penetration of the majority opinion, which here corresponds to a decision to vote or vaccinate.} 
Lastly, the model is shown to compare favorably to a \tcb{na\"ive} linear prediction method.
\end{abstract}


\section{Introduction}

A wide range of dynamical models have been proposed to capture social interactions and dynamics of opinions in large groups of people. The tutorials \cite{proskurnikov2017tutorial}, \cite{proskurnikov2018tutorial} review dominant frameworks in the field and their respective area of application. These models all base the local dynamics of interpersonal interaction on sociological principles, while seeking to avoid undue complexity in an effort to mirror the simplicity of the physical laws of nature. As a consequence, rich analytic results have been derived and numerous extensions to different settings have been proposed in the literature; recent works include \cite{Bizyaeva2023, altafini13antagonistic, yu11directed}. While some of the basic models have been empirically validated in limited settings (e.g. \cite{friedkin15community}) the same does not hold for the many offshoots and more complex variations. This highlights the importance of studies to verify that the proposed dynamics resemble real social systems. Large-scale application to existing data is performed in \cite{devia22analyze}, \cite{bernardo21paris}, and is a valuable alternative to infer the validity of dynamics without resorting to more resource-intensive tailored studies. 

The most basic type of dynamical model for opinion formation traces its roots back to the seminal work \cite{french56formal}, which laid the groundwork for linear averaging between agents interacting according to a network structure. One major shortcoming of such models is the restriction to dynamics that converge to a consensus of all agents, while many real social processes never reach such a state. Instead, agents frequently end up in a state of permanent disagreement, or \textit{dissensus}. Attempts to remedy this have been made, notably with the extension of Friedkin and Johnsen \cite{friedkin1990social}, which preserve the linear averaging dynamics while in general converging to dissensus. One strength of this extension is the empirical validity, shown for small groups as reported in \cite{friedkin15community}. While successful, the model fails to capture several phenomena observed in opinion formation, such as polarization of public opinion. The bounded confidence model in \cite{deffuant00mixing} shows another approach to preserving linear averaging dynamics while using nonlinearity to limit the interaction of distant agents. This instead typically results in several disconnected clusters reaching local consensus.

Indeed, no single model can be expected to accurately describe the myriad different situations in which social interactions take place. The present work focuses on the formation of group opinion on decisions between mutually exclusive alternatives using the model first proposed in \cite{bakovic24multipolar}. Similar dynamics with multipolar choice are found in \cite{Bizyaeva2023}. However, the model proposed therein investigates the use of bifurcations analysis to explain how a deadlock is broken in situations where a rapid group decision is required. We instead treat cases where the final distribution is spread out across the spectrum of different options, and the decision is individual. Relevant scenarios include, for example, general elections or media consumption. In this work, we apply the model to public data gathered by the governmental agency Statistics Sweden on two different issues: the rate of electoral participation and the coverage of \mbox{COVID-19} vaccination \cite{SCB2022}. 

A defining feature of the model is the allocation of \textit{bias} to each agent, which indicates their predisposition towards the possible options. As demonstrated in \cite{bakovic24multipolar}, the distribution of bias terms and the structure of the underlying graph together determine the outcome of the model; the degree of polarization varies greatly when agents of similar bias are grouped together or dispersed across the graph. We argue that this makes the model especially suitable for analysis of issues where the bias of each agent naturally correlates with the topological properties of the social network. Examples include the prototypical political divide between rural and urban populations and the concentration of individuals belonging to a certain social class in a specific area as a consequence of gentrification or housing segregation. For the studied datasets in particular, the place of residence is closely correlated to factors such as social class and immigration background, which in turn strongly influence the measured variables \cite{SCB2022}. \tcb{In this work, we will consider data on the willingness of the population in different regions to participate in the COVID-19 vaccination program, which is in turn used to predict electoral participation as proxy for the individual trust toward governmental institutions.}

In our study, the graph determining the structure of interaction is based solely on geographical data, using the local regional divisions defined in \cite{SCB2022}, but we generate individual connections according to the Watts-Strogatz model \cite{Watts1998}. Recent works \cite{pagan19networks}, \cite{pagan21meritocratic} shed light on how social networks arise based on data from online social media. While a more empirically motivated graph model would allow for greater interpretation of the results on a granular level, we suggest that the approximate approach in this preliminary work is sufficient to draw conclusions from the aggregate results. The model manages to capture the local polarization of opinion on the target issue (vaccination coverage), correctly identifying segregated areas with high levels of predisposition that deviate from the majority. On a population level, the overall distribution of opinions closely matches the measured data, showing that the proposed dynamics do indeed perform realistically given appropriate selection of the model parameters.

The remainder of the paper is organized as follows. Section II states the dynamical model used and recapitulates its basic properties. Section III details the methodology used in the application to the data, including the choice of graph and interpretation of the model parameters. The structure of the datasets is reviewed in Section IV. Finally, Section V presents and discusses the results, followed by a summary of the drawn conclusions in Section VI.

\section{Opinion-Dynamical model}

The present work seeks to validate the dynamical model proposed in \cite{bakovic24multipolar} for opinion formation \tcb{for a particular case}. We first give a brief summary of the dynamics in question and some key features of the model. The reader is referred to the original article for a more detailed description and examples illustrating the dynamics. 

Consider a number of agents, each represented by a state vector ${\mathbf{x}^i\in\mathbb{R}^k_+}$, at all times constrained to the unit simplex of dimension~${k-1}$, i.e. ${||\mathbf{x}^i||_1 = 1}$. This is to be interpreted as a distribution representing the preference of agent~$i$ over~$k$ mutually exclusive options. Additionally, each agent is assigned an individual bias vector $\mathbf{r}^i\in\mathbb{R}^k_+$. Each entry represents the predisposition of agent $i$ toward the corresponding option. The agents are connected by an underlying communication graph, with $\mathcal{N}^i$ denoting the neighborhood of agent $i$. Let ${R^i=\text{diag}(\mathbf{r}^i)}$. At each time step the dynamics propagate according to
\begin{equation}\label{eq:dyn}
    \mathbf{x}^i(t+1) = \frac{\mathbf{x}^i(t) + R^i\sum_{j\in\mathcal{N}^i}\mathbf{x}^j(t)}{||\mathbf{x}^i(t) + R^i\sum_{j\in\mathcal{N}^i}\mathbf{x}^j(t)||_1}.
\end{equation} 

While this system is nonlinear and analysis of its convergence properties is far from trivial, numerical analysis show that the model generally converges to a single attractive fixed point. The location of the fixed point is determined by the combination of biases~$\mathbf{r}^i$ and the structure of the graph, \tcb{largely} independently of the initial conditions~$\mathbf{x}^i(0)$. Exceptions from this rule are initializations 
${\mathbf{x}^i(0) = e^\ell \;\; \forall i}$,
where $e^\ell$ is the $\ell$th unit vector, as these points are trivially fixed under iteration of~\eqref{eq:dyn}. The attractive fixed point is either one of the vertices, i.e. a consensus on one of the options, or a dissensus, where agents are spread out across the interior of the simplex with no agent in a vertex~\cite[Corollary 5]{bakovic24multipolar}. Typically, a consensus is the result of homogeneous biases $\mathbf{r}^i$, while the interior dissensus is fixed and attractive given sufficiently heterogeneous biases.


\section{Methodology for data validation}

In this section, we propose a general methodology for using the multipolar opinion dynamical model presented in \cite{bakovic24multipolar} to \tcb{investigate} relationships between variables \tcb{that are assumed} to be both correlated and influenced by the topology of the interaction graph. The predictor variable is encoded in the bias $R^i$, and the outcome variable deduced from the final state $\mathbf{x}^i$ of agents in the model. Depending on the type of variable, encoding the bias and deducing the outcome can become more or less involved. In this paper, we deal with the more straightforward case of two variables which each represent a mutually exclusive choice or stance, meaning $\mathbf{x}^i$ is in the 1-dimensional unit simplex and $R^i \in \mathbb R^{2 \times 2}$. An example of this would be using a simplified stance on the question of abortion (measured as pro or contra) to model a simplified stance on gun ownership (pro or contra). 

\tcb{When real-valued variables or categorical variables with more than two categories are considered, the modeling process becomes more complex.} For example, one could consider using data on median household income as bias to predict an opinion on tax cuts. The median income data would have to be categorized and appropriately linked to probable correlated opinions on tax cuts, which possibly includes more detail than a plain yes or no choice. Our methodology supports such scenarios, but with a trade-off in complexity and the number of methodological choices which possibly require a justification from social science. 

\tcb{The dataset must provide either individual-level measurements or regionally aggregated values for the variables in question, along with sufficient spatial information to recover the geographic structure. This spatial component is essential, as the dynamics of the model are influenced by how biases are distributed across space. As demonstrated in \cite{bakovic24multipolar}, minority opinions vanish unless the opinion holders reside in tightly connected clusters in the graph.} 

\subsection{Choice of network graph}

The examination of various real networks such as online social networks, power grids and transportation networks has highlighted some key properties of the underlying interaction graphs~\cite{Watts1998}. Namely, they exhibit high clustering and low average distances. In the context of social networks, the first concept would imply that people with mutual friends are likely to be friends themselves. The second would imply that any two individuals are linked by a short chain of mutual acquaintances -- often referred to as the six degrees of separation~\cite{Watts2002}.

One of the random graph models capturing these phenomena, and the one used in our analysis, is the \tcb{well-known }Watts-Strogatz model. \tcb{Their two-step process was proposed in~\cite{Watts1998}. Initially, the graph is} a regular ring lattice with neighborhoods of size~${k \in \mathbb N}$. \tcb{During the rewiring stage, every edge is reassigned} uniformly at random with probability~${p \in [0, 1]}$. This means that topological proximity decides the likelihood of connection. \tcb{As noted later} in Section V.B, a consequence of this construction is that local neighborhoods are randomly connected, where in reality the connections on the local level are likely very structured, depending on social factors other than geographic location and \tcb{such real-world aspects are especially challenging to model. One limitation of the Watts-Strogatz model is the failure to generate hubs - nodes with an extremely high number of connections. This can be accounted for by using a preferential attachment model instead.}

\tcb{Due to the specifics of this process, nodes which start off closer together will with high probability be closely connected in the final graph. This property can be used to ensure that geographically close regions are also closely connected in the graph. The encoding of geographical information is then achieved by giving agents from neighboring regions sequential indices.}

\subsection{Bias assignment}
Supposing that the variable chosen for the bias represents a choice between two mutually exclusive options~$a$ and~$b$, agents are split into two groups $\mathcal V_a, \mathcal V_b$ such that ${\mathcal V_a \cap \mathcal V_b = \emptyset}$ and ${\mathcal V_a \cup \mathcal V_b = \mathcal V}$. 
Interpreting~$a$ as the choice tied to the first opinion index, every agent $i$ from group~$\mathcal V_a$ is assigned a bias~${\mathbf{r}^i= \begin{bmatrix} 1-\epsilon & \epsilon \end{bmatrix}}$, while every agent $j$ from group~$\mathcal V_b$ is assigned a bias~${\mathbf{r}^j= \begin{bmatrix} \epsilon & 1-\epsilon \end{bmatrix}}$, where~$\epsilon$ \tcb{is small, modeling the assumption that agents have a strong predisposition for one alternative or choice, based on the data, but would have had some probability of being swayed another way.}

In the absence of evidence of the basis of individual's decisions, we take choice as a proxy for a strong bias, hence the small~$\epsilon$. 
The ratio between the sizes of~$\mathcal V_a$ and~$\mathcal V_b$ should be consistent with the dataset, not just on the level of a global average but within each geographical subdivision. This means that every graph neighborhood assigned to a real geographical subdivision should bare the same ratio between agents who chose~$a$ and those who chose~$b$. One should also take care in connecting geographical subdivisions to graph neighborhoods, making sure that closely connected subdivisions are represented by clusters -- graph neighborhoods with high connectivity. In case of a Watts-Strogatz graph, these will be all of the consecutive sequences of agents.

\subsection{On opinions}

As noted in \cite{bakovic24multipolar}, starting opinions do not affect the final distribution in the general case of heterogeneous bias, as long as they are initialized strictly in the interior of the simple. The initial opinions can therefore be set at will, we typically use~$\mathbf x^i(0) = {[0.5 \; 0.5]}$.

The final opinions are used to infer the outcome variable. One way to do so would be to decide on a threshold. For two mutually exclusive options, every opinion $\mathbf x_i$ lies on a 1-simplex, so as long as the threshold is larger than $0.5$ this results in a unique outcome. This approach raises an additional question of where to put the threshold. To avoid introducing an additional parameter, the average of the opinion in the index corresponding to an affirmative answer can be interpreted as the percentage of agents choosing the option. This aligns with a probabilistic view of the model, where an opinion on the simplex can be viewed as a distribution describing the probability of choosing each possible option.

\begin{figure}[!h]
    \centering
    \includegraphics[width=\linewidth]{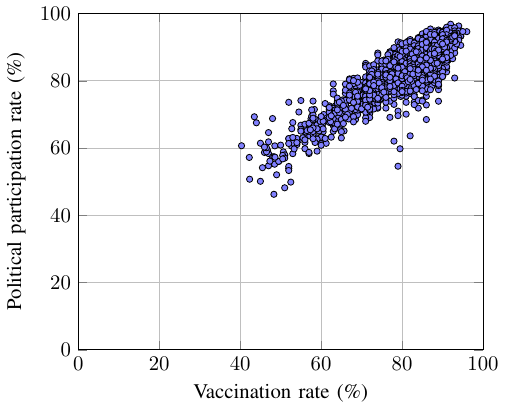}
    \caption{The dataset~\cite{SCB2022} used for our analysis, dots represent regional statistical units}
    \label{fig:scb}
\end{figure}

\section{Dataset review}

In 2022, Statistics Sweden published a report exploring the relationship between vaccination coverage and political participation~\cite{SCB2022}. Specifically, the percentage of people vaccinated against COVID-19 with at least two doses was compared to the percentage of people participating in the 2019 local elections. The data was collected on the level of demographic statistical regions (DeSO), the smallest geographical partitions of Sweden, which divide the country into 5984 units. Other divisions include 3363  regional statistical areas (RegSO) and 290 municipalities. The relationship between the variables was analyzed on all three levels and a positive correlation was found. A scatter plot of the dataset can be seen in 	\tcb{Fig.}~\ref{fig:scb}. Furthermore, the dataset displayed an interesting spatial component. One finding was that regions with low vaccine coverage are also those with low political participation rates, whilst regions with high vaccination rates also display high voter turnout. Furthermore, regions with low vaccine coverage seem to be clustered, mostly on the outskirts of the three largest cities but also in certain areas of the countryside. 


What makes this dataset particularly applicable to our methodology is the fact that both of the measured variables are mutually exclusive choices. Namely, the decision to take two doses of the vaccine and the decision to vote in the elections. On a general level, both actions can be viewed as a willingness to participate in collective societal events, whether it be a common effort to overcome a pandemic or a common effort to choose the next leadership. It is thus reasonable to believe that such variables would be correlated. In this particular case, each of the variables could be chosen as the predictor or the outcome. Lastly, the data also has a geographical component making it suitable for the study of spatial correlations, and their role in final opinion distributions. An important result from the previous paper~\cite{bakovic24multipolar} was that spatially correlated biases $\mathbf{r}^i$ lead to opinion segregation.

\subsection{Numerical validation setup}
This subsection describes the steps for applying the model to the SCB dataset.
All simulations were performed using a Watts-Strogatz graph of just under 80000 nodes, randomly generated with the parameters ${p=0.2}$ and ${k=8}$. (The particular network size was chosen with regard to computation time, and could otherwise be much larger with a longer running time of the opinion dynamics algorithm as the trade off.) With our chosen neighborhood size and rewiring probability, the network exhibits an average path length of $5.86$ making it closely resemble real social networks. 

Graph neighborhoods were coupled to RegSO partitions in the following way. A list of all partitions and their populations was sorted by municipality. All partitions within a municipality were then given consecutive neighborhoods of the graph. As explained earlier, consecutive sequences of agents are well-connected with each other. Giving partitions within a municipality consecutive neighborhoods ensures that they also have a slightly higher connectivity internally, whilst different municipalities connect with each other almost uniformly. The sizes of each graph neighborhood were determined proportional to the fraction of the corresponding partition in relation to the whole population.

The vaccination rate was then used to distribute biases following the process described in the methodology section \tcb{using $\epsilon = 0.05$}, with pro-vaccination agents~${i \in \mathcal V_a}$ being assigned a bias of~${\mathbf{r}^i=[0.95 \, 0.05]}$ and anti-vaccination agents~${j \in \mathcal V_b}$ being assigned~${\mathbf{r}^i=[0.05 \; 0.95]}$. \tcb{The model displays a certain degree of robustness and small variations in the $\epsilon$-parameter produce similar opinion distributions.}The political participation rate for each geographical unit was then inferred from the average of the first opinion index in the corresponding unit of the graph.

\section{Results and discussion}

\begin{figure}
    \centering
    \includegraphics[width=\linewidth]{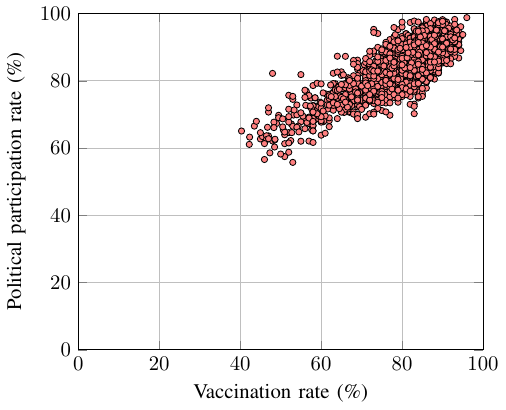}
    \caption{Model output in the base case. The political participation rate is determined by the average of the first opinion index in each graph neighborhood corresponding to a regional statistical unit. Comparison to the measured data in \tcb{Fig.} \ref{fig:scb} shows a close resemblance of the overall shape.}
    \label{fig:modelout1}
\end{figure}


\begin{figure}
    \centering
    \begin{tikzpicture}
    \begin{axis}[
        ybar,
        ylabel= \# of RegSO units,
        xlabel= Political participation (\%),
        ymin=0,
        ymax=400,
        xmin=40,
        xmax=100,
        xtick={40, 50, 60, 70, 80, 90, 100},
        legend pos=north west,
        legend cell align=left,
        area legend,
        grid=major
    ]
        \addplot+ [
                    hist={data min=0,data max=100,bins=80}, 
                    opacity=0.5,
                    color=black,
                    fill=blue!60
                  ]
            table [y index=1] 
            {scb_dataset.txt};
        \addplot+ [
                    hist={data min=0,data max=100,bins=80}, 
                    opacity=0.5,
                    color=black,
                    fill=red!60
                  ]
            table [y index=1] 
            {A_val.txt};
        \legend{voting data,model output}
    \end{axis} 
    \end{tikzpicture}
    \caption{Comparing the histograms, we can see that the model skews slightly to the right. The close adherence to the measured outcome supports the validity of the model on an aggregate level. However, comparison with the more granular data presented in \tcb{Fig.} \ref{fig:skane} reveals that this does not hold locally. As discussed below, this local inconsistency is likely due to the approximate nature of the graph on the local level.} 
    \label{fig:modelhist1}
\end{figure}
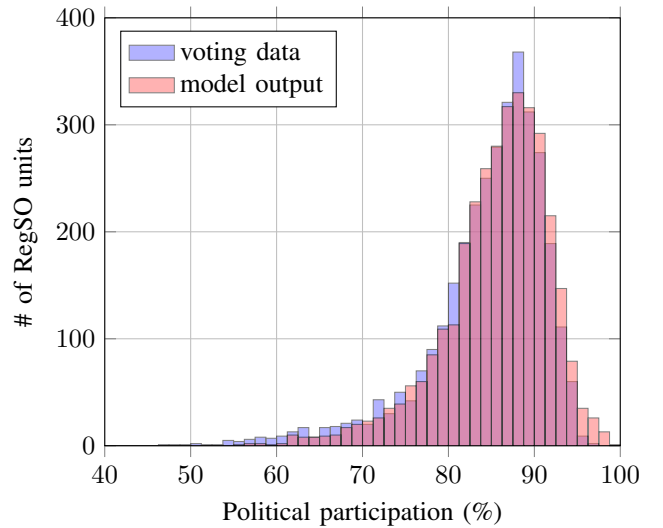

\subsection{Analysis of the output}

A scatter plot of the simulated output can be seen in \tcb{Fig.}~\ref{fig:modelout1}. The model succeeds in capturing the overall shape of the original dataset seen in \tcb{Fig.}~\ref{fig:scb}. Namely, we can see a linear trend in the data along with a variance that increases in both directions away from the point where the vaccination rate is $65\%$. For lower values of the vaccination rate, the model predicts higher political participation rates than those exhibited by the dataset. The histogram in \tcb{Fig.}~\ref{fig:modelhist1} shows once again that the model succeeds in capturing the overall shape of the target distribution, the main difference being a slight skew to the right of the output distribution. It is important to note that this analysis relies on a correlation between the examined variables. Since this is assumed, the task of predicting the output variable in itself is of trivial importance. The goal, however, is to show that the model can indeed reproduce realistic results when parameters are appropriately chosen.

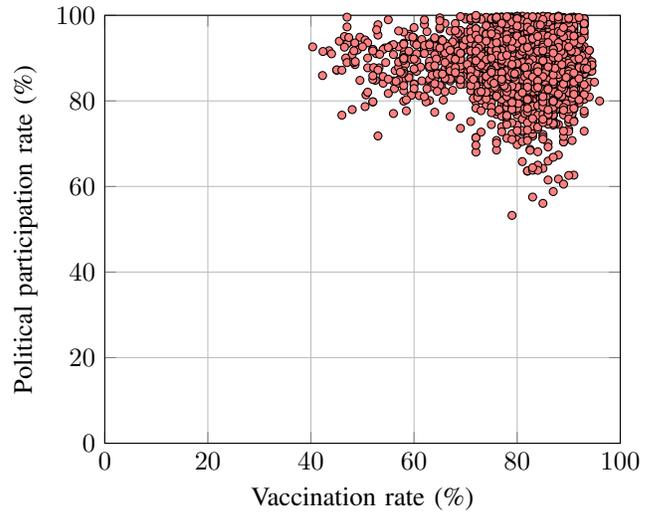
\begin{figure}
    \centering
    \begin{tikzpicture}
    \begin{axis}[
                    xlabel=Vaccination rate (\%),
                    ylabel=Political participation rate (\%),
                    xmin=0,
                    xmax=100,
                    ymin=0,
                    ymax=100,
                    grid=major
                ]
        \addplot+ [
                    only marks, 
                    mark size=1.5pt,
                    mark options={fill=red!50},
                    color = black
                  ] table 
            {R_val.txt};
    \end{axis}
    \end{tikzpicture}
    \caption{When the same total amount of pro and anti-vaccination agents are distributed uniformly amongst the graph, the results stop resembling the target dataset. This highlights the importance of topology when using the multipolar model -- or in other words, the fact that spatially correlated biases $\mathbf{r}^i$ strongly affect the final opinion distribution.}
    \label{fig:rand}
\end{figure}

The spatial correlation of biases in small regions is important for the result. If the same overall number of pro-vaccination and anti-vaccination agents is distributed along the graph without matching the regional percentages, the final result is qualitatively radically different as can be seen from 	\tcb{Fig.}~\ref{fig:rand}. This clearly shows that a detailed realistic model of the social graph is crucial to the quality of the result, and that basic properties such as connectedness and average degree are not sufficient to explain the observed phenomena. 

\begin{figure*}[!ht]
    \centering
    \begin{minipage}{0.4\textwidth}
        \centering
        \includegraphics[width=0.8\textwidth]{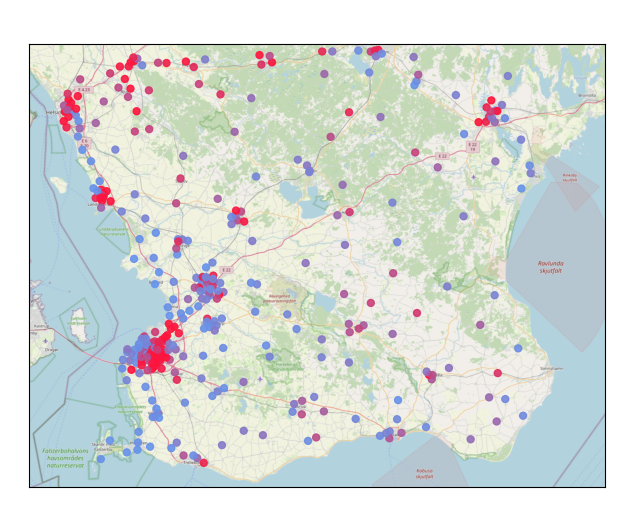} 
        \subcaption{Voting participation in the SCB dataset}
        \label{fig:spstart}
    \end{minipage}\hfill
    \begin{minipage}{0.2\textwidth}
        \centering
        \vspace{-10pt}
        \includegraphics[width=0.4\textwidth]{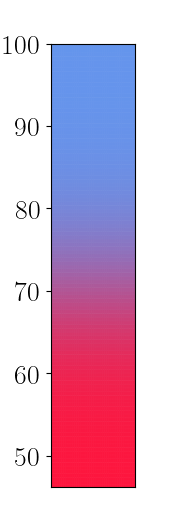} 
    \end{minipage}\hfill
    \begin{minipage}{0.4\textwidth}
        \centering
        \includegraphics[width=0.8\textwidth]{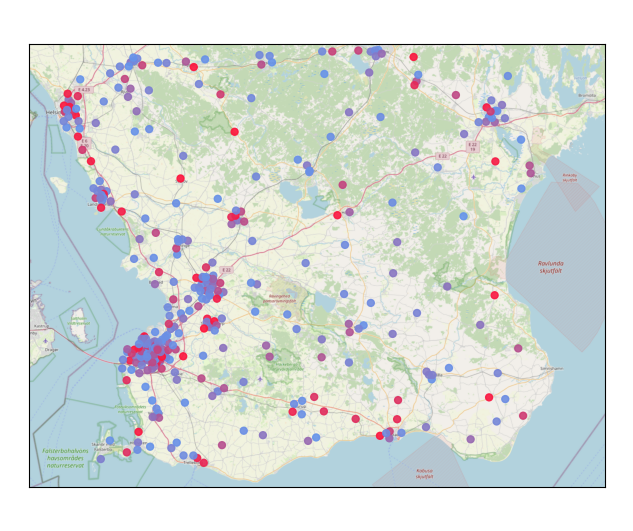} 
        \subcaption{Model output, interpreted as participation}
        \label{fig:spint}
    \end{minipage}
    \caption{The geographical comparison is most evident when looking at the region's largest city, Malmö. While the model correctly predicts that this area will experience polarization of opinions instead of consensus, the measured polarization borders are much sharper than they appear in the model output. One reason is that our network only captures geographical topology. In reality, individuals connect over many different networks at their workplace, university or local neighborhood - meaning that geographical distance is insufficient for a full description.}
    \label{fig:skane}
\end{figure*}

 \tcb{Fig.}~\ref{fig:skane} displays a map of South Sweden, the dots representing centroids of regional statistical areas. Comparing the data on participation to the model output, we see that our approach manages to capture the phenomenon of opinion polarization in larger cities. Such areas can be completely identified from the model output as the cities of Malmö, Lund, Landskrona, Helsingborg and Kristianstad. The polarization measured by the dataset often has sharper borders than the output of our model, as can be seen looking at the largest city in the region, Malmö. One explanation is that our network is purely geographical, the topology is formed taking into account only physical distance. In reality, we observe that individuals connect through many different network topologies which can rely on class, race or ethnicity to name a few. \tcb{Assuming connection probabilities based on socioeconomic factors available in other datasets could possibly generate a more socioeconomically accurate neighborhood model. This subject is part of an ongoing work.}

It is worth noting that established models of opinion dynamics do not usually exhibit such modes of convergence. Consensus models would give the result equal to a weighted average of the starting opinions. Using the bounded confidence model would result in several disjoint peaks in the distribution (consensus inside localized neighborhoods). Either of those would fail to capture the nature of the dataset. 


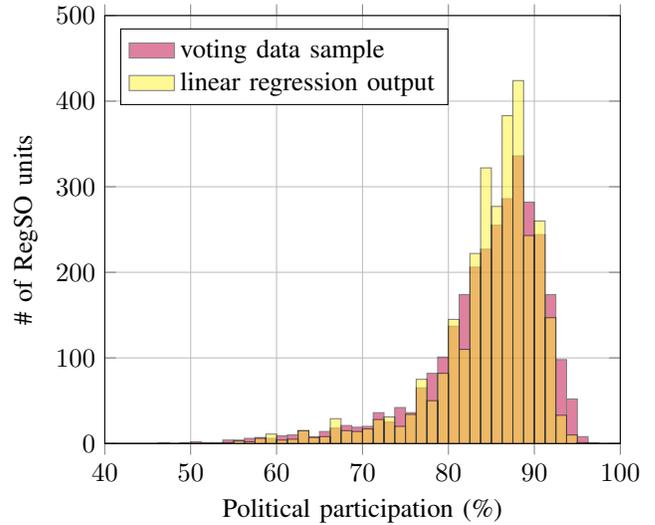
\begin{figure}[!h]
    \centering
    \begin{tikzpicture}
    \begin{axis}[
        ybar,
        ylabel= \# of RegSO units,
        xlabel= Political participation (\%),
        ymin=0,
        ymax=500,
        xmin=40,
        xmax=100,
        xtick={40, 50, 60, 70, 80, 90, 100},
        legend pos=north west,
        legend cell align=left,
        area legend,
        grid=major
    ]
        \addplot+ [
                    hist={data min=0,data max=100,bins=80}, 
                    opacity=0.5,
                    color=black,
                    fill=purple
                  ]
            table [y index=0] 
            {L_val.txt};
        \addplot+ [
                    hist={data min=0,data max=100,bins=80}, 
                    opacity=0.5,
                    color=black,
                    fill=yellow!90
                  ]
            table [y index=1] 
            {L_val.txt};
        \legend{voting data sample,linear regression output}
    \end{axis} 
    \end{tikzpicture}
    \caption{A comparison of the histograms of the prediction achieved by linear regression and the target dataset reveals that, while it does achieve a low RMSE, the distribution has many isolated peaks and seems to be overall concentrated more towards the mean of the dataset.}
    \label{fig:linear}
\end{figure}

\subsection{Prediction with linear regression}

When linear relationships between variables are suspected, social science often turns to regression models to capture the underlying correlation. In this subsection we illustrate a scenario where linear regression might be used and compare it to the results from our dynamics. Suppose that the vaccination rates in the population are known and the political participation rates are the target. If a linear relationship between the two is suspected, a part of the population could be polled to determine the relationship on a data sample, and the rest of the values could then be inferred using the resulting regression.

We partition the dataset into a training set of 300 data points, and an evaluation set of 3034 data points. The resulting inference on the target set is showcased in \tcb{Fig.}~\ref{fig:linear}. Whilst the linear regression has a lower mean square error than our model ($0.011$ on average compared to $0.0025$), we argue that the contribution of our model is capturing the overall shape of the data better than a line would and showcasing the effect of spatial correlation in the data.

Examining its histogram, we see that linear regression produces many isolated peaks which do not appear in the target set. A possible explanation is that the underlying process exhibits some degree of averaging and dissipation of opinions which the regression cannot capture.

\section{Conclusion}
We have proposed a framework for using a multipolar model to explore segregation in correlated variables in large populations of individuals. The method is general and allows for variables of arbitrary dimension, as long as proper care is taken in encoding the bias and inferring the target variable from the final state of the model. When tested on a real data set, the model captures the main features of the target variable and achieves a root mean square error of the same order of magnitude as linear regression. We have demonstrated the importance of network topology to the problem by a comparison to the output achieved by assigning biased agents to random graph neighborhoods.

One direction for future work is furthering this specific investigation of vaccination coverage in segregated communities by including variables such as level of education. An analysis of different topologies could also prove useful, for example making the graph connections encode, e.g., class in addition to geography. Validation in more complex scenarios is also desirable, especially moving beyond the binary choices examined in the present work. This presents challenges both in finding appropriate data and in the encoding of biases. \tcb{A study of corresponding datasets in other countries could prove itself fruitful in that regard.} 

\tcb{The insight that a mixing of biases leads to higher vaccine coverage opens up for intervention strategies. The general question of influencing the final opinion distribution is also a subject of ongoing work. } Lastly, theoretical analysis of the model is ongoing, in hope that intuition gained by simulations such as these can be transferred to provable properties of the nonlinear opinion dynamical system.

\bibliographystyle{ieeetr}
\bibliography{references}

\end{document}